\documentclass[conference]{IEEEtran}
\IEEEoverridecommandlockouts

\usepackage{hhline}
\usepackage{bm}
\usepackage{color}
\usepackage{longtable}	
\usepackage{multicol}
\usepackage{float}
\usepackage{multirow}
\usepackage{enumerate}
\usepackage{cite}
\usepackage{amsmath,amssymb,amsfonts}
\usepackage{algorithmic}
\usepackage{enumitem}
\usepackage{textcomp}
\usepackage{diagbox}
\usepackage{xcolor}
\usepackage{graphicx}
\usepackage{nicefrac}

\ifCLASSOPTIONcompsoc
\usepackage[caption=false,font=normalsize,labelfon
t=sf,textfont=sf]{subfig}
\else
\usepackage[caption=false,font=footnotesize]{subfi
g}
\fi

\def\BibTeX{{\rm B\kern-.05em{\sc i\kern-.025em b}\kern-.08em
    T\kern-.1667em\lower.7ex\hbox{E}\kern-.125emX}}

\begin{document}

\IEEEoverridecommandlockouts

\title{Online PMU-Based Method for Estimating Dynamic Load Parameters in Ambient Conditions\\
\thanks{This work is supported by the Natural Sciences and Engineering Research Council (NSERC) under Discovery Grant NSERC RGPIN-2016-04570 and the Fonds de Recherche du Qu\'ebec-Nature et technologies under Grant FRQ-NT PR-253686.
}
}

 \author{
 \IEEEauthorblockN{Georgia Pierrou, Xiaozhe Wang}
 \IEEEauthorblockA{Department of Electrical and Computer Engineering, McGill University, Montreal, QC H3A 0E9, Canada\\
 georgia.pierrou@mail.mcgill.ca,  xiaozhe.wang2@mcgill.ca}
 }


\maketitle

\begin{abstract}
This paper proposes a novel purely measurement-based method for estimating dynamic load parameters in near real-time when stochastic load fluctuations are present. By leveraging on the regression theorem for the Ornstein-Uhlenbeck process, the proposed method can provide highly accurate estimation without requiring any information about the power system model. Furthermore, by exploiting recursion, an online algorithm is developed to estimate the dynamic load parameters in near real-time. Simulation studies on the IEEE 39-Bus system demonstrate the accuracy of the model-free method, its robustness against measurement noise, as well as its satisfactory performance in tracking real-time changes of the load parameters.
\end{abstract}


\section{Introduction}
Dynamic load modeling plays an important role in power system stability. Traditionally, load dynamics have been considered as the driving parameters towards voltage instability \cite{Cutsem}. Inaccurate load models may lead to misconceptions regarding the stability limits and control decisions \cite{Makarov96}. To ensure reliability in power system operation and to avoid costly system controller design, it is essential that the load dynamic behavior is accurately captured and load characteristics are properly identified. However, load identification remains a challenging issue due to the large number, the complexity and the time variability of the different load components as well as the uncertainty brought about by poor measurements and customer behavior \cite{Bai09}.

To address load parameter estimation, different approaches have been used in the literature. According to the component-based  approach \cite{Price88}, load models are built up based on the true inventory of typical loads. Nevertheless, such information may not be always available. On the other hand, measurement-based methods \cite{Bai09}, \cite{Lin93}--\cite{ Wangxz:2017} can provide direct information about the load parameters through systematic monitoring \cite{Lin93}. The weighted least squares method has been applied in \cite{Lin93} to derive the load model. In \cite{Guo12}, the nonlinear least square (NLS) technique is used to identify the parameters of a dynamic recovery load in the presence of voltage changes due to load tap changer operation. An improved NLS method is proposed in \cite{Zhang17}, however, an initial guess 
on parameter values 
is required for the estimation. A hybrid learning technique that combines the Genetic Algorithm and the Levenberg-Marquard algorithm for load parameter estimation has been proposed in \cite{Bai09}. However, the implementation of the aforementioned optimization-based methods leads to high computational burden, making the real-time application prohibitive \cite{Rouhani16}.

Limited work has been done on real-time measurement-based load parameter estimation. An unscented Kalman Filter  for online estimation of load parameter values is implemented in \cite{Rouhani16}.
Multiple model estimation algorithm is used in \cite{Najafabadi13} for a composite load parameter estimation. However, a priori information has to be assumed to determine the possible ranges of the parameter values. Also, enough system excitation is required to estimate the dominant parameters, whereas estimation results may not be responsive enough for less sensitive parameters, reducing the overall accuracy.  
In \cite{Wangxz:2017}, a hybrid measurement-based method has been developed to estimate dynamic load parameters by combining the load modeling and the statistical properties of states' measurements. Yet, the proposed algorithm requires the static load characteristics and the variance of the stochastic load fluctuations to be known.

In this paper, we 
propose a novel real-time measurement-based method for load parameter estimation based on the regression theorem for the Ornstein-Uhlenbeck process. The main contributions of the paper are as follows:
\begin{enumerate}
    \item  The proposed approach is completely model-free without 
    the need of knowing any information about the static and statistical characteristic of loads (as apposed to \cite{Wangxz:2017}), or parameter values of the network model.
    \item The proposed method is computationally efficient and can be implemented online by exploiting recursion. 
    \item The variation of dynamic load parameters can be accurately captured and identified in near real-time.
\end{enumerate}

The remainder of the paper is organized as follows: Section \ref{1} introduces the dynamic load model. Section \ref{2} reviews the regression theorem for the multivariate Ornstein-Uhlenbeck process and describes the proposed methodology for online load parameter estimation. Section \ref{3} provides a complete numerical study on the IEEE 39-bus system that demonstrates the effectiveness of the method. Section \ref{4} presents the conclusions and perspectives.

\section{The Stochastic Dynamic Power System Model}
\label{1}
Although load dynamics are the main focus, generator dynamics are also modelled in the simulation for a more realistic set-up. In particular, we number the load buses as $k=1,2,..m$ and the generator buses as $i=m+1,...,N$. The generators are
described by the classical swing equations as follows:
\vspace{-3pt}
\begin{eqnarray}
\label{eq:dyngen}
\dot{\delta}_i &=& \omega_{i} \\
M_{i} \dot{\omega}_i &=& P_{mi}-P_{Gi}(\delta_{i}, \theta_{i}, V_{i}) -D_i\omega_{i}
\end{eqnarray}
\vspace{-20pt}
\begin{eqnarray}
\label{eq:generation}
P_{Gi}(\delta_{i}, \theta_{i}, V_{i}) = \sum_{j=1}^{N} |V_i||V_j|(G_{ij}\cos \theta_{ij}+B_{ij}\sin \theta_{ij})\\
Q_{Gi}(\delta_{i}, \theta_{i}, V_{i}) = \sum_{j=1}^{N} |V_i||V_j|(G_{ij}\sin \theta_{ij}-B_{ij}\cos \theta_{ij}) \label{eq:generation2}%
\end{eqnarray}

\noindent where

\begin{tabular}{cp{0.4\textwidth}}
  $\delta_{i}$ & generator rotor angle \\
$\omega_{i}$ & generator rotor angular speed \\
$M_{i}$ & inertia constant \\
$P_{mi}$ & mechanical power input \\
$P_{Gi}$ & active power injection\\
$Q_{Gi}$ & reactive power injection\\
$D_{i}$ & damping coefficient\\
$N$ & total number of buses\\
$|V_{i}|$ & $i$-th bus voltage magnitude\\
$G_{ij}$ & conductance between bus $i$ and $j$ \\
 $B_{ij}$ & susceptance between bus $i$ and $j$ \\
 $\theta_{ij}$ & difference between $i$-th and $j$-th bus voltage angles
\end{tabular}\\

As for the dynamic loads, we apply the dynamic load model proposed in \cite{Nguyen16}, which has been shown to be able to naturally characterize a wide range of loads (e.g. induction motors, thermostatic loads). For the $k$-th load, 
we have:
\vspace{-3pt}
\begin{eqnarray}
\label{eq:dynload}
\dot{g}_k &=& -\frac{1}{\tau_{gk}}(P_k-P_k^s) \label{eq:conductance}\\
\dot{b}_k &=& -\frac{1}{\tau_{bk}}(Q_k-Q_k^s)\label{eq:susceptance}
\end{eqnarray}
\vspace{-20pt}
\begin{eqnarray}
\label{eq:loaddemand}
P_k = g_kV_k^2 = \sum_{j=1}^{N} |V_k||V_j|(-G_{kj}\cos \theta_{kj}-B_{kj}\sin \theta_{kj})\\
Q_k = b_kV_k^2 = \sum_{j=1}^{N} |V_k||V_j|(-G_{kj}\sin \theta_{kj}+B_{kj}\cos \theta_{kj})
\end{eqnarray}

\noindent where

\begin{tabular}{cp{0.4\textwidth}}
  $g_{k}$ & effective load conductance \\
   $b_{k}$ & effective load susceptance \\
 $\tau_{gk}$ & active load power time constant \\
 $\tau_{bk}$ & reactive load power time constant \\
$P_k$ & active load power demand\\
$Q_k$ & reactive load power demand\\
  $P_k^s$ & steady-state active load power demand\\
 $Q_k^s$ & steady-state reactive load power demand\\
\end{tabular}\\
\newline
Note that the instant real and reactive power consumed by the dynamic load can be described by the effective conductance and susceptance as shown in (\ref{eq:conductance})-(\ref{eq:susceptance}). Also,  the authors of \cite{Nguyen16} show that distinction between different types of loads mainly lies in the parameter values of time constants $\tau_{gk}$ and $\tau_{bk}$.  





To describe the natural random variation of loads, we follow similar approaches as in \cite{Wangxz:2017}, \cite{Nwankpa93}, \cite{Pierrou19} to add Gaussian noises to the steady-state power:
\vspace{-3pt}
\begin{eqnarray}
\label{eq:stochdynload1}
\dot{g}_k &=& -\frac{1}{\tau_{gk}}[P_k-P_k^s(1+\sigma_{k}^p\xi_{k}^p)] \\
\dot{b}_k &=& -\frac{1}{\tau_{bk}}[Q_k-Q_k^s(1+\sigma_{k}^q\xi_{k}^q)]
\label{eq:stochdynload2}
\end{eqnarray}
where $\sigma_{k}^p, \sigma_{k}^q$ describe the intensity of the stochastic load variation and $\int_{0}^{t} \xi_{k}^p(u) du, \int_{0}^{t} \xi_{k}^q(u) du$ are Brownian motions. It is worth noting that stochastic load variations have been commonly modeled to follow Gaussian distribution \cite{Singh10}, since the stochastic fluctuation is the aggregate behavior of thousands of individual devices acting independently.

By linearizing  (\ref{eq:stochdynload1})-(\ref{eq:stochdynload2}) and representing them in the vector form, we have:
\vspace{-4pt}
\begin{eqnarray}
\begin{bmatrix} \dot{\bm{g}} \\ \dot{\bm{b}}
\end{bmatrix} &=& \begin{bmatrix} -T_{g}^{-1}\frac{\partial \bm{P}}{\partial \bm{g}} & -T_{g}^{-1}\frac{\partial \bm{P}}{\partial \bm{b}}\\ -T_{b}^{-1}\frac{\partial \bm{Q}}{\partial \bm{g}}&  -T_{b}^{-1}\frac{\partial \bm{Q}}{\partial \bm{b}}  \end{bmatrix}
\begin{bmatrix} \bm{g} \\ \bm{b} \end{bmatrix} \nonumber \\ &+& \begin{bmatrix} T_{g}^{-1}P^s\Sigma^p  & \bm{0} \\ \bm{0} &  T_{b}^{-1}Q^s\Sigma^q  \end{bmatrix} \begin{bmatrix}
\bm{\xi^p} \\ \bm{\xi^q}  \end{bmatrix} \nonumber \\
\label{eq:matrixload}
\end{eqnarray}

where

\begin{tabular}{ l l }
 & \\
  $\bm{g} = \begin{bmatrix} g_{1}, ..., g_{m} \end{bmatrix}^T$,  & $\bm{b} = \begin{bmatrix} b_{1}, ..., b_{m} \end{bmatrix}^T$, \\
  $T_g = $ diag $\begin{bmatrix} \tau_{g1}, ..., \tau_{gm} \end{bmatrix}$, & $T_b = $ diag $\begin{bmatrix} \tau_{b1}, ..., \tau_{bm} \end{bmatrix}$, \\
  $\bm{P}= \begin{bmatrix} P_{1}, ..., P_{m} \end{bmatrix}^T$, &
  $\bm{Q}= \begin{bmatrix} Q_{1}, ..., Q_{m} \end{bmatrix}^T$,\\
  $P^s= $ diag $\begin{bmatrix} P_{1}^s, ..., P_{m}^s \end{bmatrix}$, &  $Q^s= $ diag $ \begin{bmatrix} Q_{1}^s, ..., Q_{m}^s \end{bmatrix}$, \\
  $\Sigma_p = $ diag $\begin{bmatrix} \sigma_{1}^p, ..., \sigma_{m}^p \end{bmatrix}$, & $\Sigma_q = $ diag $\begin{bmatrix} \sigma_{1}^q, ..., \sigma_{m}^q \end{bmatrix}$,\\
  $\bm{\xi^p} = \begin{bmatrix} \xi_{1}^p, ..., \xi_{m}^p \end{bmatrix}^T$, &
  $\bm{\xi^q} = \begin{bmatrix} \xi_{1}^q, ..., \xi_{m}^q \end{bmatrix}^T$
\end{tabular}\\

Let $\bm{x}= \begin{bmatrix} \bm{g}, \bm{b} \end{bmatrix}^T $, $A = \begin{bmatrix} -T_{g}^{-1}\frac{\partial \bm{P}}{\partial \bm{g}} & -T_{g}^{-1}\frac{\partial \bm{P}}{\partial \bm{b}}\\ -T_{b}^{-1}\frac{\partial \bm{Q}}{\partial \bm{g}}&  -T_{b}^{-1}\frac{\partial \bm{Q}}{\partial \bm{b}}  \end{bmatrix}$, $ B = \begin{bmatrix} T_{g}^{-1}P^s\Sigma^p  & \bm{0} \\ \bm{0} &  T_{b}^{-1}Q^s\Sigma^q  \end{bmatrix}$, $\xi=\begin{bmatrix} \bm{\xi^p}, \bm{\xi^q}  \end{bmatrix}^T$, then (\ref{eq:matrixload}) can be written in the following compact form:
\begin{equation}
    \label{eq:OUprocess}
    \dot{\bm{x}}=A\bm{x}+B\bm{\xi}
\end{equation}
which is a vector Ornstein-Uhlenbeck process 
\cite{Pierrou:TCAS19}.
Particularly, $P_k=g_k V_k^2$, $Q_k=b_kV_k^2$ and thus:
\begin{eqnarray}
    \label{eq:dP/dg}
    \frac{\partial P_k}{\partial g_j}=\left\{
\begin{array}{cc}
     V_{k}^2+2g_jV_k \frac{\partial V_k}{\partial g_j}\approx V_k^2 \hspace{12pt} \textrm{if}\quad j=k\\ \quad \quad
     2g_jV_k \frac{\partial V_k}{\partial g_j}\approx 0  \hspace{24pt}   \textrm{if}\quad j\ne k\\
\end{array}
\right.
\\
    \label{eq:dQ/db}
    \frac{\partial Q_k}{\partial b_j}=\left\{
\begin{array}{cc}
     V_{k}^2+2b_jV_k \frac{\partial V_k}{\partial b_j}\approx V_k^2 \hspace{4pt} \textrm{if}\quad j=k\\ \quad \quad
     2b_jV_k \frac{\partial V_k}{\partial b_j}\approx 0 \hspace{16pt} \textrm{if}\quad j\ne k
\end{array}
\right.
\end{eqnarray}
\begin{equation}
 \label{eq:dP/db}
    \frac{\partial P_k}{\partial b_j}= 2g_jV_k \frac{\partial V_k}{\partial b_j}  \approx 0
\end{equation}
\begin{equation}
\label{eq:dQ/dg}
    \frac{\partial Q_k}{\partial g_j}= 2b_jV_k \frac{\partial V_k}{\partial g_j} \approx 0
\end{equation}
as $\Delta V_{k} \approx 0$ in ambient conditions.

\noindent 
Therefore, we have the following approximation for matrix $A$:
\begin{equation}
     \label{eq:matrixA}
     A \approx \begin{bmatrix}
-T_g^{-1}V^2 & \bm{0} \\
\bm{0} & -T_b^{-1}V^2
\end{bmatrix} = \begin{bmatrix}
A_{T_{g}} & \bm{0} \\
\bm{0} & A_{T_{b}}
\end{bmatrix}
\end{equation}

\noindent where
$V= \textrm{diag} \begin{bmatrix}
V_1, ..., V_m
\end{bmatrix}, A_{T_g}=-T_g^{-1}V^2$ and $A_{T_b}=-T_b^{-1}V^2$.

\section{The Proposed PMU-Based Method for Estimating Dynamic Load Parameters}
\label{2}

\subsection{The Theoretical Basis of the Method}
As shown in (\ref{eq:OUprocess}), the stochastic dynamic load equations can be represented as a vector Ornstein-Uhlenbeck process. The stationary covariance matrix $C_{\bm{xx}}$ is described by:
\vspace{-2pt}
\begin{equation}
\label{eq:covariance}
C_{\bm{xx}} = \big \langle [\bm{x}(t)-\mu_{\bm{x}}][\bm{x}(t)-\mu_{\bm{x}}]^T \big  \rangle = \begin{bmatrix}
C_{\bm{gg}} & C_{\bm{gb}} \\
C_{\bm{bg}} & C_{\bm{bb}}
\end{bmatrix}
\end{equation}
where $\mu_{\bm{x}}$ is the mean of the process. The $\tau$-lag autocorrelation matrix is:
\vspace{-2pt}
\begin{equation}
\label{eq:tlag_corelation}
G(\tau) = \big \langle [\bm{x}(t+\tau)-\mu_{\bm{x}}][\bm{x}(t)-\mu_{\bm{x}}]^T \big  \rangle
\end{equation}

According to the \textit{regression theorem} of the vector Ornstein-Uhlenbeck process \cite{Gardiner}, if the system matrix $A$ is stable, that is typically satisfied in the normal operating state of a power system, the $\tau$-lag autocorrelation matrix satisfies:
\begin{equation}
\label{eq:regressiontheorem}
\frac{d}{d\tau} [G(\tau)] = - AG(\tau)
\end{equation}
Hence, the dynamic system state matrix $A$ can be obtained from:
\begin{equation}
\label{eq:matrixA_regressiontheorem}
A=\frac{1}{\tau}\log\begin{bmatrix}G(\tau)C^{-1}\end{bmatrix}
\end{equation}
Equation (\ref{eq:matrixA_regressiontheorem}) ingeniously relates the statistical properties of PMU measurements 
with the physical model of power systems. We will leverage on (\ref{eq:matrixA_regressiontheorem}) to develop a novel measurement-based method to estimate $A$, and further to estimate  $T_g$ and $T_b$ from (\ref{eq:matrixA}). 
It is worth noting that 
the proposed method is purely measurement-based and requires no other information concerning the system model to effectively estimate matrix $A$, $T_g$ and $T_b$. Specifically, the proposed method requires no knowledge about the static characteristics of the dynamic loads ($P_k^s$ and $Q_k^s$ in (\ref{eq:stochdynload1})-(\ref{eq:stochdynload2})) and the intensities of stochastic load variations ($\sigma_k^p$ and $\sigma_k^q$ in (\ref{eq:stochdynload1})-(\ref{eq:stochdynload2})), which, conversely, have to be known in \cite{Wangxz:2017}.

\subsection{Estimating the Load Parameters from PMU measurements}
\label{proposedalgorithm}
In practice, the sample mean $\bm{\bar{x}}$, the sample covariance matrix $\hat{C}$ and the sample $\tau$-lag correlation matrix $\hat{G}$ can be estimated from a finite number of PMU measurements. Firstly, the available PMU measurements are used to estimate the sample mean $\bar{V} = \textrm{diag} \begin{bmatrix}
\bar{V}_1, ..., \bar{V}_m
\end{bmatrix}$ and $\bm{x}_{i}= \begin{bmatrix} \bm{g}_{i}, \bm{b}_{i} \end{bmatrix}^T, i=1,...,n$ as below:
\vspace{-3pt}
\begin{eqnarray}
\label{eq:Vbar}
\bar{V}_{k} &=& \frac{1}{n}\sum_{i=1}^{n} V_{ki}\\
\label{eq:estimatedg}
g_{ki} &=&\operatorname{Re}{\{\frac{I_{ki}}{V_{ki}}}\} \\
\label{eq:estimatedb}
b_{ki} &=& \operatorname{Im}{\{\frac{I_{ki}}{V_{ki}}}\}
\end{eqnarray}
where $n$ is the number of samples and $I_{ki}, V_{ki}$ are the $i$-th current and voltage phasor measurements at bus $k$.
Then, $\bm{\bar{x}}$, $\hat{C}$ and $\hat{G}$ are calculated as follows:
\vspace{-1pt}
\begin{equation}
\label{eq:xbar}
\bar{\bm{x}}= \frac{1}{n}\sum_{i=1}^{n} \bm{x}_{i}
\end{equation}
\begin{equation}
\label{eq:Chat}
\hat{C}= \frac{1}{n-1}(F-\bar{\bm{x}}\bm{1}_{n}^T)(F-\bar{\bm{x}}\bm{1}_{n}^T)^T
\end{equation}
\vspace{-4pt}
\begin{equation}
\label{eq:tlaghat}
\hat{G}(\Delta t)= \frac{1}{n-1}(F_{1+\kappa:n}-\bar{\bm{x}}\bm{1}_{n-\kappa}^T)(F_{1:n-\kappa}-\bar{\bm{x}}\bm{1}_{n-\kappa}^T)
\end{equation}
where
$F=\begin{bmatrix} \bm{x}_{1}, ...,\bm{x}_{n} \end{bmatrix}$ is a $m \times n$ matrix with the states' measurements, $\bm{1}_{n}$ is a $n \times 1$ vector of ones, $\Delta t$ is the time lag, $F_{i:j}$ denotes the submatrix of $F$ from the $i$-th to the $j$-th column, $\kappa$ is the number of samples that correspond to the selected time lag. The estimated dynamic system state matrix $\hat{A}$ can be obtained as:
\vspace{-2pt}
\begin{equation}
    \label{eq:Ahat}
    \hat{A}=\frac{1}{\Delta t}\log \begin{bmatrix} \hat{G}(\Delta t)\hat{C}^{-1}\end{bmatrix}\approx
    \begin{bmatrix}A_{UL}&0\\0&A_{LR}\end{bmatrix}
\end{equation}
where $A_{UL}$ is a diagonal matrix, whose diagonal components are equal to those of the upper left submatrix of $\hat{A}$. Similarly $A_{LR}$ is a diagonal matrix, whose diagonal components are equal to those of the lower right submatrix of $\hat{A}$.
Therefore, according to (\ref{eq:matrixA}), we can estimate ${A} _{T{_g}}$ and ${A} _{T{_b}}$ as follows:
\vspace{-2pt}
\begin{eqnarray}
    \hat{A} _{T{_g}}&=&A_{UL}\label{eq:A_UL}\\
    \hat{A} _{T{_b}}&=&A_{LR}\label{eq:A_LR}
\end{eqnarray}
The load parameters $T_{g}$ and $T_{b}$ can also be estimated:  
\vspace{-1pt}
\begin{eqnarray}
\label{estimateTg}
\hat{T}_{g} &=& -\bar{V}^2 \hat{A}^{-1} _{T{_g}}\\
\label{estimateTb}
\hat{T}_{b} &=& -\bar{V}^2 \hat{A}^{-1} _{T{_b}}
\end{eqnarray}

\subsection{The Online Recursive Estimation Algorithm}
\label{onlinealgorithm}
For online implementation, it is desirable to conduct the estimation recursively for the sake of computational efficiency. 
Once the statistics (i.e., the sample mean, the sample covariance matrix and the sample $\tau$-lag aucorrelation matrix) are estimated from a pre-defined length of PMU data, they can be updated recursively.
Assuming that PMUs are deployed at the substations where the (aggregated) loads of interest are connected, then the following algorithm can estimate the load parameters $T_{g}$ and $T_{b}$ online from PMU data:

\begin{enumerate}[label=\textbf{Step \arabic*.},ref=Step \arabic*., leftmargin=35pt]
\item Given PMU data of size $n$ from pre-defined window length, estimate the sample mean $\bar{V}$ (\ref{eq:Vbar}) and calculate $\bm{g}$ and $\bm{b}$ using (\ref{eq:estimatedg})-(\ref{eq:estimatedb}).
\item For $j=0$, compute the initial values of $\bm{\bar{x}}_{0}$, $\hat{C}_{0}$ and $\hat{G}_{0}(\triangle t)$ using (\ref{eq:xbar})-(\ref{eq:tlaghat}) 
and $\hat{A}_{0}$, $\hat{T}_{g0}$ and $\hat{T}_{b0}$ using (\ref{eq:Ahat})-(\ref{estimateTb}). 
\item For $j=1,2,...$ update as follows  \cite{Sheng19}:
\begin{enumerate}[leftmargin=-0.77cm]
    \item \begin{equation}
 \begin{gathered}
\label{eq:xbar_update}
\bm{\bar{x}}_{j}=(1-\alpha)\bm{\bar{x}}_{j-1}+\alpha x_j\\
 \hat{G}_{j}(\Delta t)=(1-\alpha)[ \hat{G}_{j-1}+\alpha(\bm{x}_j-\bm{\bar{x}}_{j})(\bm{x}_{j-1}-\bm{\bar{x}}_{j-1})^T] \\
\hat{C}_{j}^{-1}=\frac{1}{1-\alpha}[\hat{C}_{j-1}^{-1}-\alpha\frac{\hat{C}_{j-1}^{-1}z_jz_j^T\hat{C}_{j-1}^{-1}}{1+\alpha z_j^T\hat{C}_{j-1}^{-1}z_j}]
\end{gathered}
\end{equation}
\noindent where $z_j=(\bm{x}_j-\bm{\bar{x}}_{j-1})$.
\item \begin{equation}
    \label{eq:Ahat_update}
    \hat{A}_{j}=\frac{1}{\Delta t}\log \begin{bmatrix} \hat{G}_{j}(\Delta t)\hat{C}_{j}^{-1} \end{bmatrix}
\end{equation}
\item Estimate 
$\hat{T}_{g_j}, \hat{T}_{b_j}$ using (\ref{eq:A_UL})-(\ref{estimateTb}).
\end{enumerate}
\end{enumerate}

Particularly, \textbf{Step 3-a}
updates the statistics recursively given the new sample, based on which the dynamic system state matrix and the time constants can be updated recursively in \textbf{Step 3-b,c}. Sherman-Morrison formula \cite{Sheng19} is used to update the inverse of $\hat{C}_j$ recursively without calculating the inverse of the matrix directly. It can be observed that the online algorithm is computationally efficient, 
since it relies on simple mathematical operations. Note that the performance of the online algorithm may be affected by the pre-defined window length as well as the
parameter $\alpha$ \cite{Sheng19}. Intuitively, the larger the pre-defined window length is, the more accurate estimation results one may achieve. 
In addition, even though $\alpha=\frac{1}{n}$ in steady state, $\alpha$ can be termed as a smoothing parameter to adjust the weight of observations during transient conditions. Hence, if a load parameter drift is detected, a larger $\alpha$ can be used to give more weight to the new samples, whereas $\alpha$ could be decreased back to $\alpha=\frac{1}{n}$ once the new steady state estimation results are reached.
\vspace{-3pt}
\section{Numerical Results}
\label{3}
The IEEE 39-Bus 10-Generator test system has been used to validate the effectiveness of the proposed method and the online recursive algorithm. Particularly, 10 stochastic dynamic loads described as (\ref{eq:stochdynload1})-(\ref{eq:stochdynload2}) with time constants varying from 0.1s to 5s have been added to the system. Gaussian stochastic load variations with a mean of zero and $\sigma_{k}^p=\sigma_{k}^q=1$ in (\ref{eq:stochdynload1})-(\ref{eq:stochdynload2}) have been considered in the load model, where $k=1,...,m$. A time lag of $\Delta t=0.2$s has been selected, i.e. $\kappa=10$ in (\ref{eq:tlaghat}) 
given that the integration step size is 0.02s. 
All simulations were implemented in PSAT-2.1.10 \cite{Milano05}.

\vspace{-3pt}
\subsection{Validation of the Proposed Method}
\label{section:500smethod}
\begin{figure}[!tb]
\vspace{-12pt}
\centering
\includegraphics[width=2.3in ,keepaspectratio=true,angle=0]{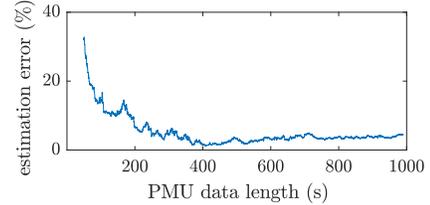}
\vspace{-10pt}
\caption{The estimation error as the sample size changes.}
\label{errorvswindowfig}
\vspace{-12pt}
\end{figure}

To validate the proposed algorithm described in Section \ref{proposedalgorithm}, 500s PMU measurements are used. The sample size of 500s is selected so that a balance between accuracy and data length is achieved. Fig. \ref{errorvswindowfig} illustrates how the estimation error, in terms of the normalized Frobenius norm $\frac{{\left\lVert A-\hat{A} \right\rVert}_{F}}{{\left\lVert A\right\rVert}_{F}}$, changes with the length of PMU data.  
The estimated time constants  
  $\hat{\tau}_{gk}, \hat{\tau}_{bk}$ and their corresponding error with respect to their true values $\tau_{gk}, \tau_{bk}$, are shown in Table \ref{table:500smethod}. It can be observed that the proposed data-driven method can well estimate the dynamic load parameters in the whole range of time constants.


\vspace{-3pt}
\subsection{Impact of Measurement Noise}

In practical application, PMU measurements may be contaminated by noise. Since the proposed method is purely measurement-based, its robustness against measurement noise should be tested. Following the approach in \cite{Zhou13}, \cite{Anagnostou18}, high noise levels with standard deviation of $10 \%$ of the largest state changes between time steps have been added to the 500s PMU measurements of $\bm{g}, \bm{b}$. Noise with standard deviation of $10^{-3}$ has also been added to the measurements of the voltage magnitudes $V$ \cite{Anagnostou18}. The estimated load time constants when measurement noise is present are shown in Table \ref{table:mn500s}. As can be observed, the accuracy of the algorithm remains satisfactory, indicating that the proposed method is robust against measurement noise. 

 \begin{table}[!tb]
\centering
  \caption{The estimated load parameters $\hat{\tau}_{gk}, \hat{\tau}_{bk}$ using 500s PMU data}\label{table:500smethod}
  \vspace{-8pt}
  \begin{tabular}{|c|c|c|c|}
\hhline{|-|-|-|-|}
\textbf{Time constant}& \textbf{Real value (s)}& \textbf{Estimated value (s)}& \textbf{Error}\\
 \hline
$\tau_{g1}$&0.1&0.1043&4.2536\%
\\
\hline
$\tau_{g2}$&0.6&0.6023&0.3825\%
\\
\hline
$\tau_{g3}$&1.1&1.2073&9.7512\%
\\
 \hline
$\tau_{g4}$&1.6&1.4821&-7.3682\%
\\
\hline
$\tau_{g5}$&2.1&2.0868&-0.6270\%
\\
\hline
$\tau_{g6}$&2.6&2.4621&-5.3030\%
\\
 \hline
$\tau_{g7}$&3.1&3.2216&3.9233\%
\\
\hline
$\tau_{g8}$&3.6&3.5127&-2.4243\%
\\
\hline
$\tau_{g9}$&4.1&3.8587&-5.8857\%
\\
 \hline
$\tau_{g10}$&4.6&4.7184&2.5729\%
\\
\hline
$\tau_{b1}$&0.5&0.4869&-2.6287\%
\\
\hline
$\tau_{b2}$&1&1.0460&4.6042\%
\\
\hline
$\tau_{b3}$&1.5&1.7968&19.7851\%
\\
 \hline
$\tau_{b4}$&2&1.9515&-2.4225\%
\\
\hline
$\tau_{b5}$&2.5&2.5685&2.7397\%
\\
\hline
$\tau_{b6}$&3&2.7340&-8.8658\%
\\
 \hline
$\tau_{b7}$&3.5&3.4069&-2.6591\%
\\
\hline
$\tau_{b8}$&4&3.8727&-3.1813\%
\\
\hline
$\tau_{b9}$&4.5&4.2547&-5.4512\%
\\
 \hline
$\tau_{b10}$&5&4.8617&-2.7668\%
\\
\hhline{|-|-|-|-|}
  \end{tabular}
  \vspace{-12pt}
\end{table}

\begin{table}[!th]
\centering
  \caption{The estimated load parameters $\hat{\tau}_{gk}, \hat{\tau}_{bk}$ using 500s PMU data with measurement noise}\label{table:mn500s}
   \vspace{-5pt}
\begin{tabular}{|c|c|c|c|}
\hhline{|-|-|-|-|}
\textbf{Time constant}& \textbf{Real value (s)}& \textbf{Estimated value (s)}& \textbf{Error}\\
 \hline
$\tau_{g1}$&0.1&0.1033&3.3358\%
\\
\hline
$\tau_{g2}$&0.6&0.5923&-1.2818\%
\\
\hline
$\tau_{g3}$&1.1&1.1937&8.5215\%
\\
 \hline
$\tau_{g4}$&1.6&1.4551&-9.0541\%
\\
\hline
$\tau_{g5}$&2.1&2.0523&-2.2738\%
\\
\hline
$\tau_{g6}$&2.6&2.4070&-7.4241\%
\\
 \hline
$\tau_{g7}$&3.1&3.1590&1.9026\%
\\
\hline
$\tau_{g8}$&3.6&3.4383&-4.4927\%
\\
\hline
$\tau_{g9}$&4.1&3.8022&-7.2628\%
\\
 \hline
$\tau_{g10}$&4.6&4.7184&2.5729\%
\\
\hline
$\tau_{b1}$&0.5&0.4813&-3.7415\%
\\
\hline
$\tau_{b2}$&1&1.0268&2.6797\%
\\
\hline
$\tau_{b3}$&1.5&1.7638&17.5836\%
\\
 \hline
$\tau_{b4}$&2&1.9135&-4.3237\%
\\
\hline
$\tau_{b5}$&2.5&2.5217&0.8667\%
\\
\hline
$\tau_{b6}$&3&2.6890&-10.3673\%
\\
 \hline
$\tau_{b7}$&3.5&3.3085&-5.4718\%
\\
\hline
$\tau_{b8}$&4&3.8109&-4.7264\%
\\
\hline
$\tau_{b9}$&4.5&4.1904&-6.8794\%
\\
 \hline
$\tau_{b10}$&5&4.8617&-2.7668\%
\\
\hhline{|-|-|-|-|}
  \end{tabular}
  \vspace{-15pt}
\end{table}


\subsection{Validation of the Online Recursive Estimation Algorithm}
\begin{figure}[!tb]
\vspace{-10pt}
\centering
\includegraphics[width=2.3in ,keepaspectratio=true,angle=0]{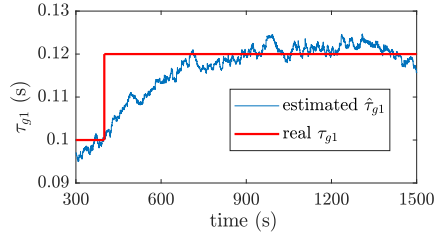}
\vspace{-10pt}
\caption{The estimated values $\hat{\tau}_{g1}$ and the actual values $\tau_{g1}$ in real-time.}
\label{esttime_change1}
\end{figure}

\begin{figure}[!tb]
\centering
\vspace{-12pt}
\includegraphics[width=2.3in ,keepaspectratio=true,angle=0]{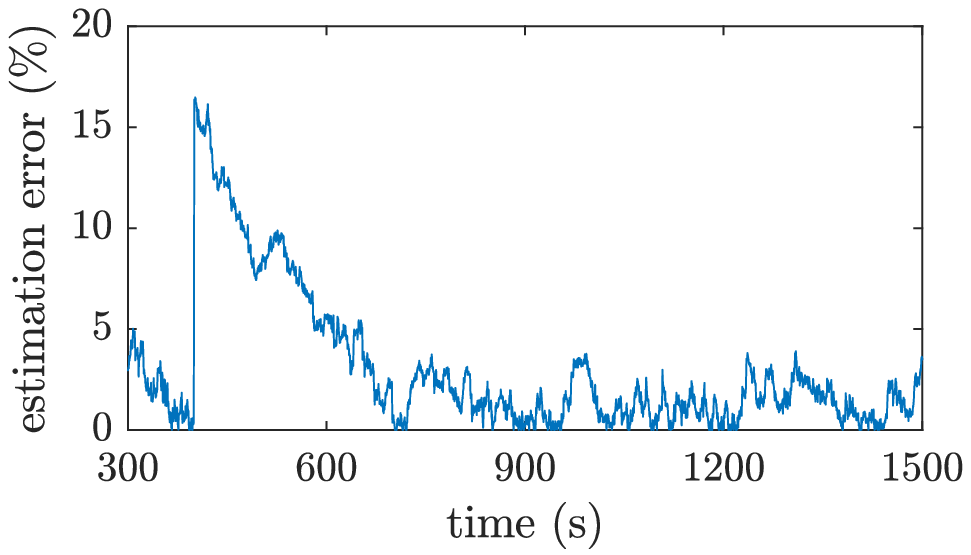}
\vspace{-5pt}
\caption{The estimation error $\frac{\left| \tau_{g1}-\hat{\tau}_{g1}\right|}{\tau_{g1}}$ in real-time.}
\label{esterror_change1}
\vspace{-12pt}
\end{figure}

\vspace{-1pt}
In this Section, we implement the online recursive estimation algorithm to evaluate its effectiveness with respect to real-time changes in the load parameters. Following a similar approach in \cite{Rouhani16}, we assume that the system experiences two scenarios of sudden load parameter changes. In the first scenario, we consider a small change where the time constant $\tau_{g1}$ is increased by 20$\%$, 
i.e. from 0.1 to 0.12, at t=400s. The results of the online estimation using a pre-defined window length of 300s PMU data and $\alpha=\frac{1}{n}$ are shown in Fig. \ref{esttime_change1} and \ref{esterror_change1}. It can be seen that the online algorithm is able to accurately detect the real-time change of the load parameter. 

\begin{figure}[!tb]
\centering
\includegraphics[width=2.3in ,keepaspectratio=true,angle=0]{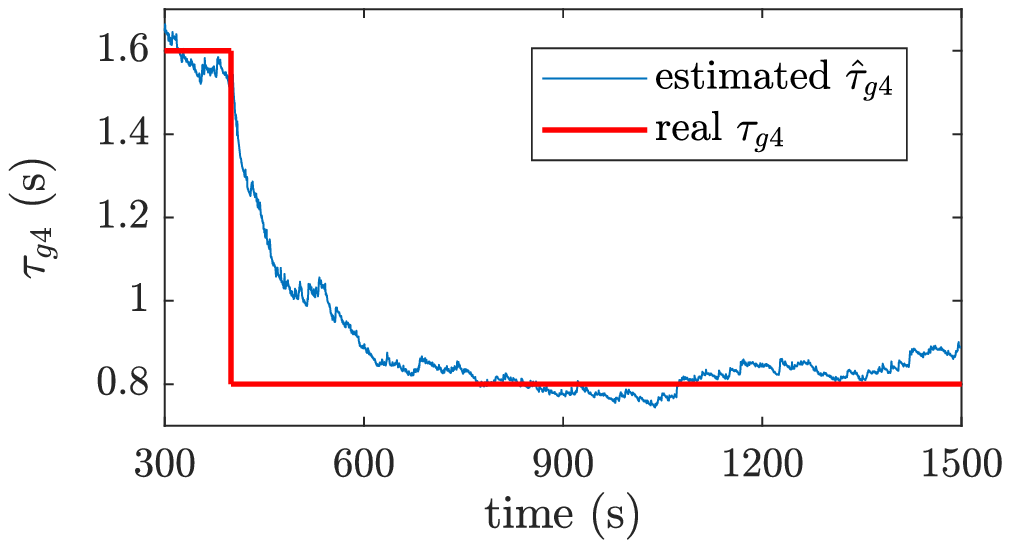}
\vspace{-8pt}
\caption{The estimated values $\hat{\tau}_{g4}$ and the actual values $\tau_{g4}$ in real-time.}
\label{esttime_change4}
\vspace{-12pt}
\end{figure}

\begin{figure}[!tb]
\centering
\includegraphics[width=2.3in ,keepaspectratio=true,angle=0]{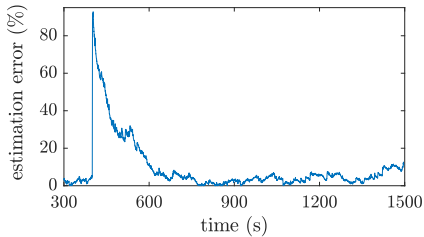}
\vspace{-4pt}
\caption{The estimation error $\frac{\left| \tau_{g4}-\hat{\tau}_{g4}\right|}{\tau_{g4}}$ in real-time.}
\label{esterror_change4}
\vspace{-12pt}
\end{figure}

In the second scenario, a more severe change is considered. 
Particularly, the time constant $\tau_{g4}$ is decreased by 50$\%$, 
i.e., from 1.6 to 0.8, at t=400s. Fig. \ref{esttime_change4} and \ref{esterror_change4} present the estimated values and the corresponding errors in real-time using a pre-defined window length of 300s PMU data and $\alpha=\frac{1}{n}$. The results show that the performance of the online algorithm remains satisfactory despite the severe time constant change.

It is worth noting that, if $5\%$ is chosen to be the threshold for the estimation error, the algorithm requires approximately 200s new PMU data to reach the new steady state estimation in both scenarios. The results are reasonable, considering the 300s pre-defined window length. As discussed in Section \ref{onlinealgorithm}, faster convergence to the new load parameter values may be reached, in case that a load parameter drift is detected and $\alpha$ is adjusted to increase the weight of the new samples accordingly. 

\vspace{-2pt}
\section{Conclusions and Perspectives}
\label{4}
\vspace{-2pt}
This paper proposes a novel measurement-based method for estimating dynamic load parameters in near real-time. By leveraging on the regression theorem  developed for the Ornstein-Uhlenbeck process, the proposed method  can  provide  accurate estimation for the dynamic load parameters without any model information. By exploiting recursion, an online algorithm has been developed to track real-time changes of the load parameters.
Numerical results in the IEEE 39-bus system demonstrate the accuracy of the method, its robustness against measurement noise and its capability of tracking real-time changes of the load parameters in near real-time.
Future work will focus on further validation with the use of real PMU data.

\vspace{-1pt}

\end{document}